\documentclass[dvipdfmx]{pasj01}
\draft

\usepackage{natbib}

\begin{document} 
\Received{2015/12/28}
\Accepted{2016/02/05}

\title{
Transiting planets as a precision clock\\
to constrain the time variation of the gravitational constant
}
\author{
  Kento Masuda\altaffilmark{1} and
  Yasushi Suto\altaffilmark{1,2}
}
\altaffiltext{1}{Department of Physics, The University of Tokyo,
  Tokyo 113-0033, Japan}
\altaffiltext{2}{Research Center for the Early Universe, 
 School of Science, The University of Tokyo, Tokyo 113-0033, Japan}
\email{masuda@utap.phys.s.u-tokyo.ac.jp}


\maketitle

\begin{abstract}
Analysis of transit times in exoplanetary systems accurately provides an
instantaneous orbital period, $P(t)$, of their member planets. A
long-term monitoring of those transiting planetary systems puts limits
on the variability of $P(t)$, which are translated into the constraints
on the time variation of the gravitational constant $G$.  We apply this
analysis to $10$ transiting systems observed by the {\it Kepler} spacecraft,
and find that $\Delta G/G\lesssim 5\times10^{-6}$ for 2009--2013, 
or $\dot{G}/G \lesssim 10^{-6}\,\mathrm{yr}^{-1}$ if $\dot{G}$ is constant.  
While the derived limit is weaker than those from other analyses, 
it is complementary to them and can be improved by analyzing 
numerous transiting systems that are continuously monitored.
\end{abstract}

\section{Introduction \label{sec:intro}}

Are the fundamental {\it constants} in Nature really {\it constant} over
the cosmological time scale? This question has a long history in physics
and cosmology, and has been discussed intensively in different contexts.
One of the most famous examples includes the Large Number Hypothesis by
\citet{Dirac1937}, who raises a possibility of the gravitational
constant $G$ being proportional to $t^{-1}$.

Recently, \citet{Anderson2015} reported a curious oscillatory trend in
the values of $G$ measured over the last three decades, $\Delta G/G
\approx 2.4\times10^{-4} \sin[2\pi t/(5.9\,{\rm
yr})+\mathrm{const.}]$. More intriguingly, they claimed that the period
and phase of the modulation are in agreement with the variation of the
length of the day of the Earth \citep{Holme2013,Speake2014}.  The
proposed modulation, however, is unlikely to reflect the real time
variation of $G$ \citep{Anderson2015}. Indeed, the amplitude of $\Delta
G$ is shown to be inconsistent with the dynamics of the solar system
\citep{Iorio2015}.  In addition, the subsequent studies \citep[][and
added appendix of Anderson et al. 2015] {2015PhRvD..91l1101S,
2015arXiv150506725P} have shown that the trend reported by
\citet{Anderson2015} is more likely to be an artifact.  Nevertheless, it
is important and interesting to discuss and compare with other
independent constraints on $\Delta G$ on such short time scales since
most of the previous literature focused on $\Delta G$ averaged over the
cosmological time scale.

For that purpose, we consider the orbital periods of transiting
exoplanetary systems in the present paper. So far, more than
4000 candidate systems have been reported by the {\it Kepler} mission
\citep{2015ApJS..217...31M}, 
which have been monitored over several years, and
approximately 1000 of them have been confirmed to host planets.  The
orbital period $P$ of such planets can be accurately determined by their
central transit times. Furthermore, a systematic search for any time
variability of the period has been performed primarily in order to probe
the gravitational interaction between multiple planets, which is
referred to as transit timing variation
\citep[TTV,][]{2005MNRAS.359..567A, 2005Sci...307.1288H}.

The TTV analysis is conventionally used to determine the mass of planets
without radial velocity follow-up observations and/or to infer the
presence of undetected perturbers.  Instead, we attempt to put a
constraint on the time variation of $G$ from the same analysis, but
focusing on those systems that exhibit no clear TTV signature.  In
particular, {\it hot Jupiters}, planets orbiting around host stars
within a week or so, are particularly suited for constraining the
variation of $G$ on time scales of months to years.

As will be shown below, our sample yields the constraint
$\dot{G}/G \lesssim 10^{-6}\,\mathrm{yr}^{-1}$ {\it if $\dot{G}$ is
constant}, which is weaker by six orders of magnitude than those based on
the pulsar timing \citep[e.g.,][]{Williams1976,Kaspi1994,Zhu2015} and
the lunar ranging \citep[e.g.,][]{Hofmann2010}. We would like to
emphasize, however, that the conventional assumption of the constant
$\dot G$ is not general, but has been introduced just for simplicity. If
$G$ would vary periodically as was claimed by \citet{Anderson2015},
planetary systems with different orbital periods would be ideal to
search for the possible resonant effect close to the variation period of
days to years, and thus the resulting constraint on the amplitude of
$\dot G$ could be stronger depending on the expected period of the
oscillation. Therefore it would be interesting to know the current limit
from the existing systems at this point in any case.

The precise data of our Solar system also put more stringent constraints
on the variation of $G$, but they are mostly sensitive to the
variability on time scales of years, and the perturbative effect of
eight planets and other bodies need to be carefully separated, as has
been performed by \citet{Iorio2015} for instance.  On the contrary, the
result from different transiting systems can be simply added for tighter
constraints because the effect of $\Delta G$ should change the period of
any system in a coherent fashion.  For these reasons, TTVs of transiting
planetary systems offer a complementary and straightforward method to
explore the variation of $G$ on shorter time scales.

\section{Transit timing variation analysis of planetary systems\label{sec:ttv}}

In the two-body problem, motion of a planet around its host star is
exactly periodic, and so are its transits.  The transit times of a
planet in a multiple planetary system, however, sometimes deviate from
the exact periodicity due to the gravitational perturbation from other
planets in the system.  This phenomenon, known as the ``transit timing
variation'' \citep[TTV,][]{2005MNRAS.359..567A, 2005Sci...307.1288H},
has been successfully modeled in tens of multi-transiting planetary
systems discovered by {\it Kepler} to determine the mass ratios of their
member planets \citep[e.g.,][]{2010Sci...330...51H, 2011Natur.470...53L,
Masuda2014} as well as to infer the existence of non-transiting planets
\citep[e.g.,][]{2012Sci...336.1133N, Masuda2013}.  

\begin{figure}
 \begin{center}
	    \includegraphics[width=0.5\textwidth, clip]{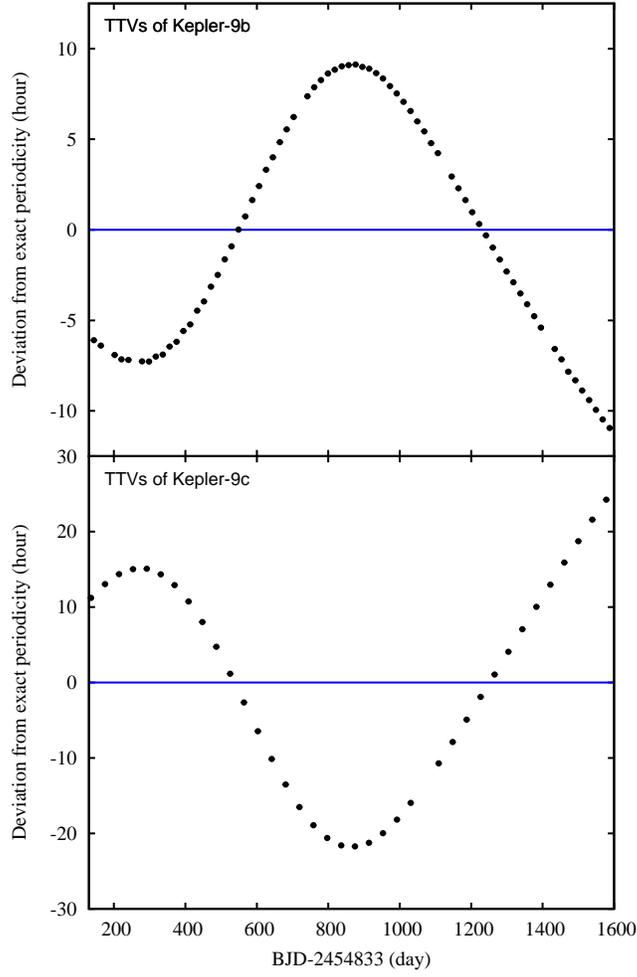}  
 \end{center}
\caption{TTVs in the Kepler-9 system, where two planets near a $2:1$
mean motion resonance (Kepler-9b with $P=19.3\,\mathrm{days}$, top;
Kepler-9c with $P=38.9\,\mathrm{days}$, bottom) transit the same host
star.  The residuals of the transit times from the linear fit are
plotted against the {\it Kepler} observational time spanning about four years.
The error of each transit time is smaller than the point size.  \label{fig:ttv}}
\end{figure}

Figure \ref{fig:ttv} displays an example of the TTV signals for the
Kepler-9 system, which is the first multiple planetary system detected
through the transit method.  It hosts two transiting planets, Kepler-9b
($0.137\,M_{\rm Jup}$) and Kepler-9c ($0.094\,M_{\rm Jup}$) with orbital
periods of $19.3\,\mathrm{days}$ and $38.9\,\mathrm{days}$, respectively
\citep{Borsato2014, 2014arXiv1403.1372D}.  
In this system,  the orbital periods of the two planets
are not exactly constant because of the strong mutual gravitational interaction
between the two in addition to the dominant gravity due to the host star
\citep[1.07\,$M_\odot$,][]{2011ApJ...727...24T}.  
This results in systematic deviations of the central
transit times with respect to the mean period as exhibited in filled circles.

Such TTV signals have often been used to constrain the system
parameters, in particular to estimate the mass of planets without radial
velocity measurement. In the following, however, we use the {\it
absence} of the TTV signal to put an upper limit on the period variation
$\Delta P$, which translates into $\Delta G$ through Kepler's third law:
\begin{equation}
\label{eq:kepler-3rd}
P = 2\pi \sqrt{\frac{a^3}{GM}} ,
\end{equation}
where $a$ is the semi-major axis and $M$ is the total mass of the two-body
system. 

Although equation (\ref{eq:kepler-3rd}) is exactly correct only when $G$
is constant, we assume that it still holds when $G$ varies adiabatically,
as we consider here. Then it leads to
\begin{equation}
\label{eq:dotP-1}
\frac{\dot{P}}{P}= \frac{3}{2}\frac{\dot{a}}{a}
- \frac{1}{2}\frac{\dot{G}}{G}
- \frac{1}{2}\frac{\dot{M}}{M}.
\end{equation}
Note that the variation of $G$, in principle, simultaneously induces non-vanishing
$\dot{a}$ and $\dot{M}$. Thus it may be more useful to
rewrite equation (\ref{eq:kepler-3rd}) in terms of the specific angular
momentum $j$ and the eccentricity $e$:
\begin{equation}
\label{eq:kepler-3rd-2}
P = \frac{2\pi j^3}{G^2M^2(1-e^2)^{3/2}} .
\end{equation}
Then we have
\begin{equation}
\label{eq:dotP-2}
\frac{\dot{P}}{P}= 3\,\frac{\dot{j}}{j}
- 2\,\frac{\dot{G}}{G}
- 2\,\frac{\dot{M}}{M}.
\end{equation}

Equation (\ref{eq:dotP-2}) simply reduces to
\begin{equation}
\label{eq:dotP-3}
\frac{\dot{P}}{P}= - 2\,\frac{\dot{G}}{G},
\end{equation}
if the orbit is circular and both the specific angular momentum and mass
are conserved under the variation of $G$
\citep[e.g.,][]{Uzan2003}. While the assumption of $\dot{M}=0$ is
perfectly justified for non-relativistic stars and planets considered
here, it is not the case for compact objects including neutron stars.

\subsection{Constraints from individual systems: Kepler-1 and Kepler-2}

\begin{figure*}
    \begin{minipage}{0.5\hsize}
    	\includegraphics[width=\textwidth, clip]{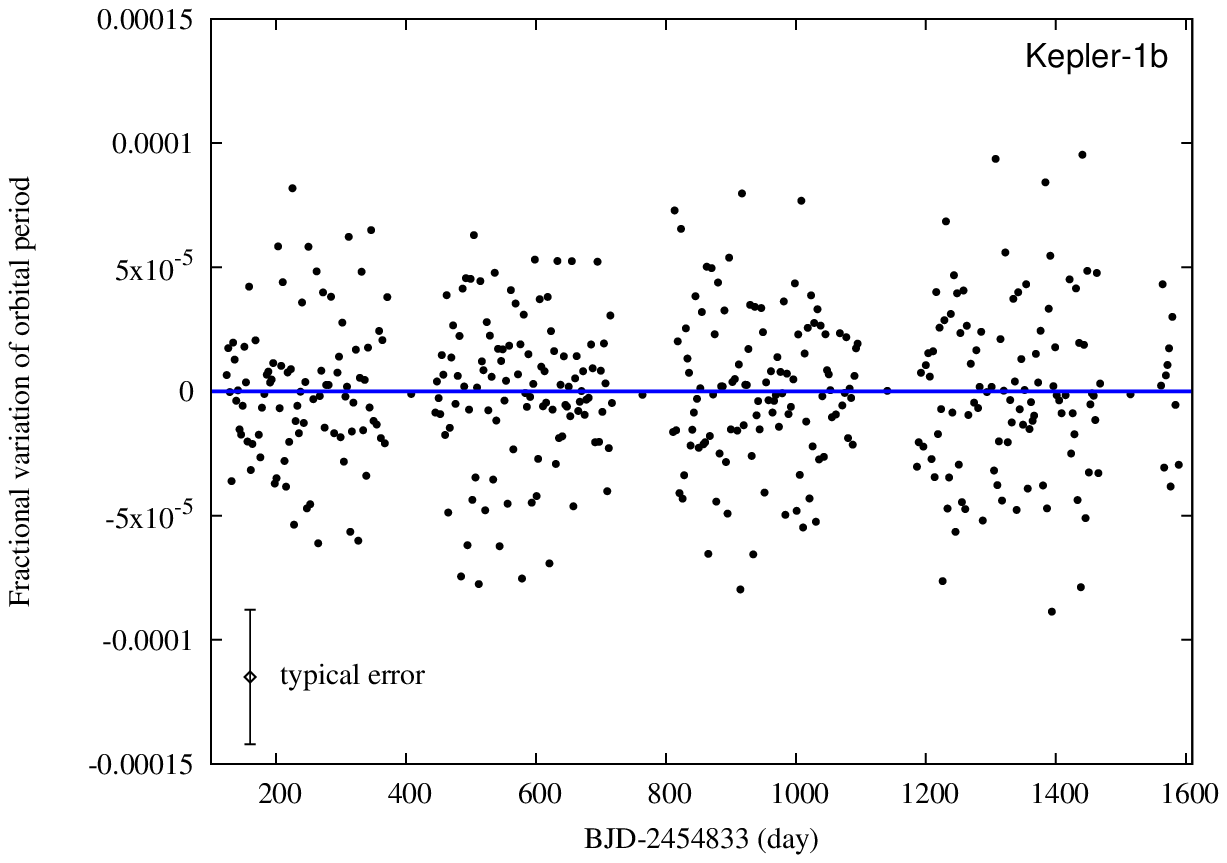}  
    \end{minipage}
    \begin{minipage}{0.5\hsize}
    	\includegraphics[width=\textwidth, clip]{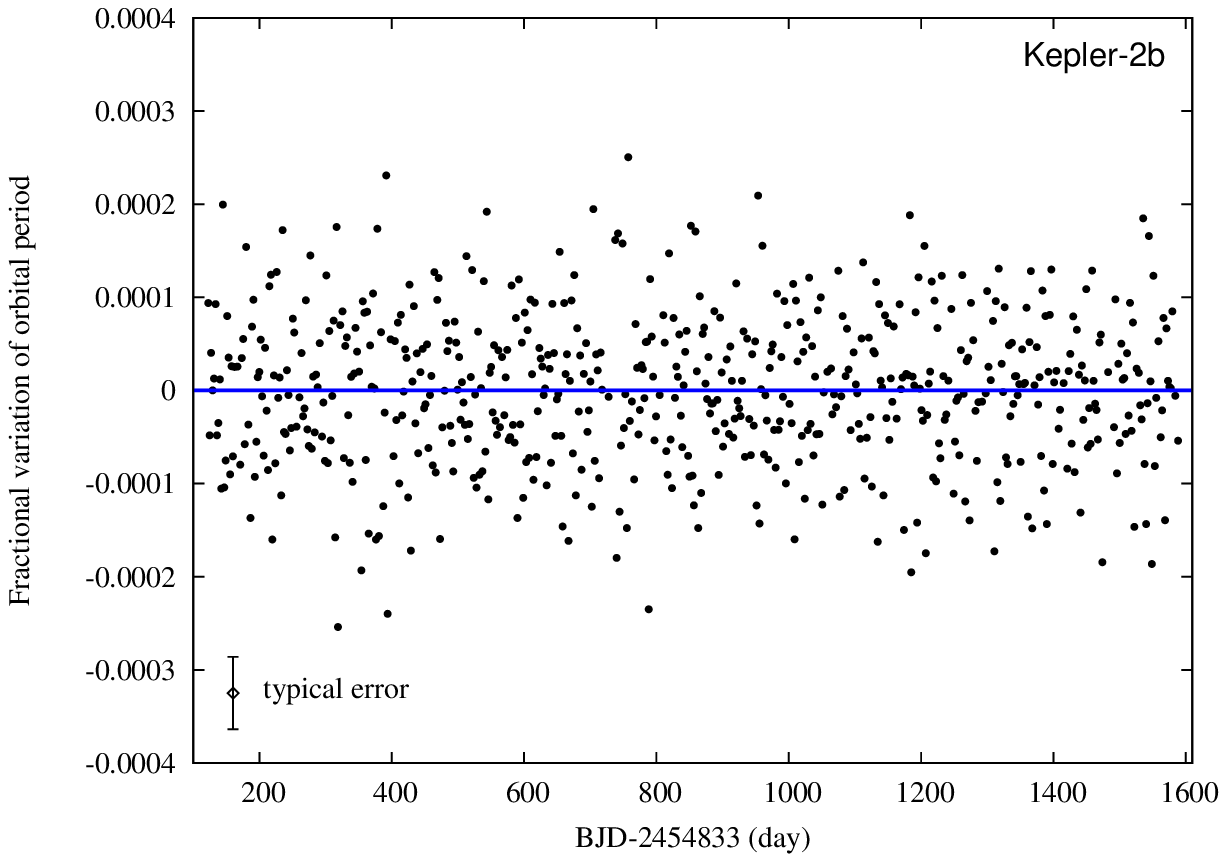}  
    \end{minipage} 
\caption{$\Delta P/P$ against $\mathrm{BJD}$ for Kepler-1b (left) and
Kepler-2b (right).  \label{fig:individual}}
\end{figure*}

Among the confirmed transiting planets observed with {\it Kepler}, we
select Kepler-1b (or TrES-2) with $1.20\,M_{\rm Jup}$ and
$2.47\,\mathrm{day}$ period, and Kepler-2b (or HAT-P-7b) with
$1.78\,M_{\rm Jup}$ and $2.20\,\mathrm{day}$ period.  The two planets have
the highest transit signal-to-noise ratio, while exhibiting no
identifiable feature of TTVs.  In Figure \ref{fig:individual}, we show
the fractional variations in the orbital periods of these planets
against the observed date.  For each planet, we follow the procedure of
\citet{2015ApJ...805...28M} to determine the central times of individual
transits $t_i$, where $i$ stands for the number of transits counted from
a fixed epoch.  Here we use only the data sampled at a short cadence
(one minute), which yield the transit times with higher precision than
the long cadence data sampled at the $30$-minute interval.  We compute
the orbital period $P_i$ between $t_i$ and $t_{i+1}$ for each $i$ and
plot $(\Delta P/P)_i \equiv(P_i - \overline{P})/\overline{P}$ against
$(t_{i+1} + t_i)/2$, where $\overline{P}$ is an average of all $P_i$.
The corresponding root-mean-square $\Delta P/P$ are $3\times10^{-5}$ and
$8\times10^{-5}$ for Kepler-1b and Kepler-2b, respectively.  These
values translate into $\Delta G/G = 1.5\times10^{-5}$ and
$4\times10^{-5}$ over four years, which are smaller than the amplitude of
the proposed variation of $G$ \citep{Anderson2015} by an order of
magnitude.

\subsection{Constraint from a statistical sample\label{ssec:statistical}}

The constraints from the above two systems almost reach the limit
of the {\it Kepler} photometry for the existing systems.  However, the
further improvement can be achieved by combining many systems in a
statistical fashion. We select 10
confirmed transiting planetary systems with the highest transit
signal-to-noise ratios: Kepler-1, Kepler-2, Kepler-13, Kepler-12,
Kepler-6, Kepler-7, Kepler-423, Kepler-17, Kepler-5, and Kepler-3.  None
of these systems exhibit any clear TTVs.  Although the transit times 
in some of these systems (e.g., Kepler-3 or alias HAT-P-11) 
are affected by the strong star-spot activities that deform the transit signals, 
we do not exclude them because our purpose here is simply to illustrate 
the advantage of combining the constraints from many independent systems.

\begin{figure*}
 \begin{center}
	\includegraphics[width=0.8\textwidth, clip]{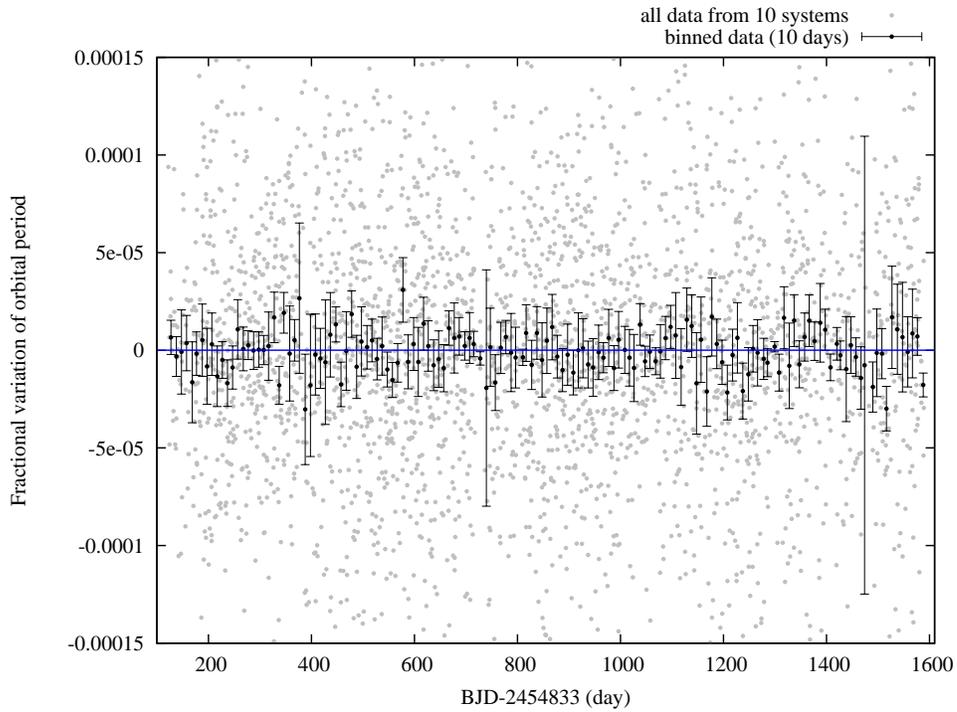} 
 \end{center}
\caption{$\Delta P/P$ against $\mathrm{BJD}$ for the 10 transiting
systems analyzed in Section \ref{ssec:statistical}. 
The gray dots are all the data points, while the black circles
with error bars are those averaged into 10-day bins.
\label{fig:statistical}}
\end{figure*}

We apply the same analysis as in the previous subsection to the above
10 systems. We plot the resulting $\Delta P/P$ as the gray dots in
Figure \ref{fig:statistical}, and their averages in $10$-day bins as the
black circles with error bars. Now the standard deviation of the binned
$\Delta P/P$ is $1\times10^{-5}$, which is a few times smaller than the
constraints from the individual systems.  Note that the choice of the
smoothing bin size is completely arbitrary at this point.  The binning
would smooth out a possible variation of $G$ less than the bin size and
the resulting $\Delta P/P$ would depend on the bin size.  Here we choose
the 10-day bin just for definiteness, and could use a different value 
if a specific model of the variation of $G$ is given.

\begin{figure*}
 \begin{center}
	\includegraphics[width=0.8\textwidth, clip]{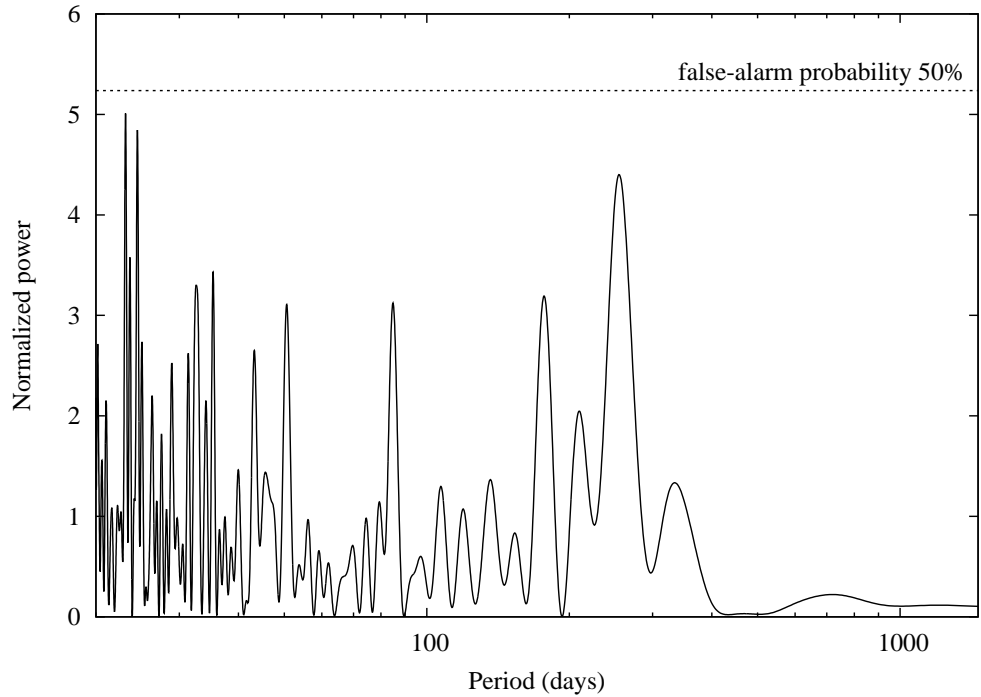}  
 \end{center}
\caption{Lomb--Scargle periodogram \citep{1982ApJ...263..835S} of the binned data in Figure \ref{fig:statistical}.
The vertical axis is normalized to the variance of the data.
Horizontal dashed line indicates the power level corresponding to the false-alarm probability 
(significance level) of $50\%$;
this is the probability that any of the peaks exceeds a given power level when the data points 
are independent Gaussians.
\label{fig:periodogram}}
\end{figure*}

The periodogram of the binned $\Delta P/P$ in Figure
\ref{fig:periodogram} delivers a rough idea of the expected constraints
for the different choice of the smoothing bin size.\footnote{ We made
use of the pyTiming module of PyAstronomy
({https://github.com/sczesla/PyAstronomy}) to compute the
periodogram.}  All of the peaks in the periodogram have the amplitudes
less than the significance level of $50\%$ (horizontal dashed line) and
are consistent with the Gaussian noise, i.e., the data exhibit no
significant modulation over the period range down to $\sim 10\,\mathrm{days}$.  A
larger bin size therefore should put a somewhat stronger constraint
on $\Delta P/P$, and hence on $\Delta G/G$.  The proper interpretation
of the constraint, however, depends on the model of $G(t)$.

\section{Conclusion and discussion \label{sec:conclusion}}

We have shown that the methodology of using exoplanets as a precision
clock provides reaonably interesting limits on the time variation of
$G$. Our current analysis finds $\Delta G/G \lesssim 5\times10^{-6}$ for 2009--2013, 
which corresponds to 
$\dot{G}/G\lesssim10^{-6}\,\mathrm{yr}^{-1}$ if $\dot{G}$ is constant. 
In contrast, constraints from Lunar Laser Ranging experiments and the binary pulsar
system PSR B1913+16 correspond to
$\dot{G}/G<8\times 10^{-12}\,\mathrm{yr}^{-1}$ \citep{1996PhRvD..53.6730W} and
$\dot{G}/G=(4\pm5)\times 10^{-12}\,\mathrm{yr}^{-1}$ \citep{Kaspi1994},
respectively.

Although our current constraint is significantly less stringent than the
binary pulsar timing result that is based on the similar principle, it
is complementary in many aspects. First, we can safely neglect $\dot{M}$
for exoplanetary systems unlike binary pulsar systems where the
self-gravitational energy significantly contributes to the total mass.
Second, we may constrain the $\dot{a}$-term independently of $\dot{P}$
by combining the precise measurement of the radial velocity
(proportional to $a/P$) with the photometric transit timing data.
Third, the timing analysis of the secondary eclipse (opposite phase of
the planetary transit in front of the host star, i.e., the occultation
of the planet by the stellar disk) of transiting planets may also
constrain the $\dot{a}$-term through the variation of the expected
arrival time difference $\approx a/c$ from the photometric data
alone. Fourth, the growing number of continuously monitored transiting
planets promises to statistically improve the constraint beyond the
value that we discussed in this paper. Incidentally, if $G$ varies
periodically with an oscillation period of days to years, 
the dynamics of the planetary system with the similar orbital period would be perturbed
in a resonant fashion. While we have not studied such a possible
resonant effect in this paper, constraints from systems with different
periods are important once the simple assumption of constant $\dot G$ is
abandoned.

Finally, we would like to emphasize that the current data for
numerous exoplanetary systems are already precise enough to 
put complementary and meaningful constraints on the variation of the
fundamental constants, as has been feasible only through the solar
system data and/or a few binary pulsars. Thus future data of the
exoplanetary systems should definitely improve the situation and might
even bring an unexpected surprise.

\begin{ack}
We thank Teruaki Suyama for calling our attention to
\cite{Anderson2015}.  K.M. is supported by JSPS Research Fellowships for
Young Scientists (No. 26-7182), and also by the Leading Graduate Course
for Frontiers of Mathematical Sciences and Physics.  Y.S. acknowledges
the support from the Grant-in Aid for Scientific Research by JSPS
(No. 24340035).
\end{ack}



\begin{thebibliography}{50}
\expandafter\ifx\csname natexlab\endcsname\relax\def\natexlab#1{#1}\fi

\bibitem[{{Agol} {et~al.}(2005){Agol}, {Steffen}, {Sari}, \&
  {Clarkson}}]{2005MNRAS.359..567A}
{Agol}, E., {Steffen}, J., {Sari}, R., \& {Clarkson}, W. 2005, \mnras, 359, 567

\bibitem[{{Anderson} {et~al.}(2015)}]{Anderson2015}
Anderson, J.~D., Schubert, G., Trimble, V., and Feldman, M.~R. 2015, EPL
110, 10002

\bibitem[{Borsato} {et~al.}(2014)]{Borsato2014}
Borsato, L., Marzari, F., Nascimbeni, V., Piotto, G., Granata, V.,
Bedin, L.~R., and Malavolta, L. 2014, \aap, 571, A38

\bibitem[{Dirac}(1937)]{Dirac1937}
Dirac, P.~A.~M. 1937, Nature, 139, 323

\bibitem[{{Dreizler} \& {Ofir}(2014)}]{2014arXiv1403.1372D}
{Dreizler}, S., \& {Ofir}, A. 2014, ArXiv e-prints, arXiv:1403.1372

\bibitem[{Hofmann} {et~al.}(2010)]{Hofmann2010}
{Hofmann}, F., M\"{u}ller, J., and Biskupek, L. 2010, Astronomy and
	      Astrophysics, 522, L5

\bibitem[{{Holman} {et~al.}(2010){Holman}, {Fabrycky}, {Ragozzine}, {Ford},
  {Steffen}, {Welsh}, {Lissauer}, {Latham}, {Marcy}, {Walkowicz}, {Batalha},
  {Jenkins}, {Rowe}, {Cochran}, {Fressin}, {Torres}, {Buchhave}, {Sasselov},
  {Borucki}, {Koch}, {Basri}, {Brown}, {Caldwell}, {Charbonneau}, {Dunham},
  {Gautier}, {Geary}, {Gilliland}, {Haas}, {Howell}, {Ciardi}, {Endl},
  {Fischer}, {F{\"u}r{\'e}sz}, {Hartman}, {Isaacson}, {Johnson}, {MacQueen},
  {Moorhead}, {Morehead}, \& {Orosz}}]{2010Sci...330...51H}
{Holman}, M.~J., {Fabrycky}, D.~C., {Ragozzine}, D., {et~al.} 2010, Science,
  330, 51

\bibitem[{{Holman} \& {Murray}(2005)}]{2005Sci...307.1288H}
{Holman}, M.~J., \& {Murray}, N.~W. 2005, Science, 307, 1288

\bibitem[{Holme} and {de Viron}(2013)]{Holme2013}
Holme R., and  de Viron O., 2013, Nature, 499, 202

\bibitem[{Iorio (2015)}]{Iorio2015}
Iorio, L. 2015, arXiv:1504.07233

\bibitem[{{Kaspi} {et~al.}(1994)}]{Kaspi1994}
{Kaspi}, V.~M., Taylor, J.~H., \& Riba, M.~F. 1994, \apj, 428, 713

\bibitem[{{Lissauer} {et~al.}(2011){Lissauer}, {Fabrycky}, {Ford}, {Borucki},
  {Fressin}, {Marcy}, {Orosz}, {Rowe}, {Torres}, {Welsh}, {Batalha}, {Bryson},
  {Buchhave}, {Caldwell}, {Carter}, {Charbonneau}, {Christiansen}, {Cochran},
  {Desert}, {Dunham}, {Fanelli}, {Fortney}, {Gautier}, {Geary}, {Gilliland},
  {Haas}, {Hall}, {Holman}, {Koch}, {Latham}, {Lopez}, {McCauliff}, {Miller},
  {Morehead}, {Quintana}, {Ragozzine}, {Sasselov}, {Short}, \&
  {Steffen}}]{2011Natur.470...53L}
{Lissauer}, J.~J., {Fabrycky}, D.~C., {Ford}, E.~B., {et~al.} 2011, \nat, 470,
  53

\bibitem[{Masuda {et~al.}(2013)}]{Masuda2013} Masuda, K., Hirano, T.,
Taruya, A., Nagasawa, M., and Suto, Y. 2013, \apj, 778, 185

\bibitem[{Masuda (2014)}]{Masuda2014}
{Masuda}, K. 2014, \apj, 783, 53

\bibitem[{{Masuda}(2015)}]{2015ApJ...805...28M}
{Masuda}, K. 2015, \apj, 805, 28

\bibitem[{{Mullally} {et~al.}(2015){Mullally}, {Coughlin}, {Thompson}, {Rowe},
  {Burke}, {Latham}, {Batalha}, {Bryson}, {Christiansen}, {Henze}, {Ofir},
  {Quarles}, {Shporer}, {Van Eylen}, {Van Laerhoven}, {Shah}, {Wolfgang},
  {Chaplin}, {Xie}, {Akeson}, {Argabright}, {Bachtell}, {Barclay}, {Borucki},
  {Caldwell}, {Campbell}, {Catanzarite}, {Cochran}, {Duren}, {Fleming},
  {Fraquelli}, {Girouard}, {Haas}, {He{\l}miniak}, {Howell}, {Huber}, {Larson},
  {Gautier}, {Jenkins}, {Li}, {Lissauer}, {McArthur}, {Miller}, {Morris},
  {Patil-Sabale}, {Plavchan}, {Putnam}, {Quintana}, {Ramirez}, {Silva Aguirre},
  {Seader}, {Smith}, {Steffen}, {Stewart}, {Stober}, {Still}, {Tenenbaum},
  {Troeltzsch}, {Twicken}, \& {Zamudio}}]{2015ApJS..217...31M}
{Mullally}, F., {Coughlin}, J.~L., {Thompson}, S.~E., {et~al.} 2015, \apjs,
  217, 31

\bibitem[{{Nesvorn{\'y}} {et~al.}(2012){Nesvorn{\'y}}, {Kipping}, {Buchhave},
  {Bakos}, {Hartman}, \& {Schmitt}}]{2012Sci...336.1133N}
{Nesvorn{\'y}}, D., {Kipping}, D.~M., {Buchhave}, L.~A., {et~al.} 2012,
  Science, 336, 1133

\bibitem[{{Pitkin}(2015)}]{2015arXiv150506725P}
{Pitkin}, M. 2015, ArXiv e-prints, arXiv:1505.06725

\bibitem[{{Scargle}(1982)}]{1982ApJ...263..835S}
{Scargle}, J.~D. 1982, \apj, 263, 835

\bibitem[{{Schlamminger} {et~al.}(2015){Schlamminger}, {Gundlach}, \&
  {Newman}}]{2015PhRvD..91l1101S}
{Schlamminger}, S., {Gundlach}, J.~H., \& {Newman}, R.~D. 2015, \prd, 91,
  121101

\bibitem[{Speake and Quinn (2014)}]{Speake2014}
Speake, C., and Quinn, T. 2014, Physics Today, 67, 27

\bibitem[{{Torres} {et~al.}(2011){Torres}, {Fressin}, {Batalha}, {Borucki},
  {Brown}, {Bryson}, {Buchhave}, {Charbonneau}, {Ciardi}, {Dunham}, {Fabrycky},
  {Ford}, {Gautier}, {Gilliland}, {Holman}, {Howell}, {Isaacson}, {Jenkins},
  {Koch}, {Latham}, {Lissauer}, {Marcy}, {Monet}, {Prsa}, {Quinn}, {Ragozzine},
  {Rowe}, {Sasselov}, {Steffen}, \& {Welsh}}]{2011ApJ...727...24T}
{Torres}, G., {Fressin}, F., {Batalha}, N.~M., {et~al.} 2011, \apj, 727, 24

\bibitem[{{Uzan}(2003)}]{Uzan2003}
{Uzan}, J.-P. 2003, Rev.Mod.Phys., 75, 403

\bibitem[{{Williams} {et~al.}(1976)}]{Williams1976}
{Williams}, P.~J., et al. 1976, Phys.Rev.Lett., 36, 551

\bibitem[{{Williams} {et~al.}(1996){Williams}, {Newhall}, \&
  {Dickey}}]{1996PhRvD..53.6730W}
{Williams}, J.~G., {Newhall}, X.~X., \& {Dickey}, J.~O. 1996, \prd, 53, 6730

\bibitem[{Zhu} {et~al.}(2015)]{Zhu2015}
{Zhu}, W. W., Stairs, I. H., Demorest, P. B., et al. 2015,
\apj, 809, 41

\end{thebibliography}

\end{document}